\documentclass[conference]{IEEEtran}
\IEEEoverridecommandlockouts
\usepackage{url}
\usepackage[hidelinks]{hyperref}
\usepackage{cite}
\usepackage{amsmath,amssymb,amsfonts}
\usepackage{algorithmic}
\usepackage{graphicx}
\usepackage{textcomp}
\usepackage{xcolor}
\usepackage{svg}
\usepackage{algorithm}
\usepackage{cancel}
\usepackage{subcaption}
\def\BibTeX{{\rm B\kern-.05em{\sc i\kern-.025em b}\kern-.08em
    T\kern-.1667em\lower.7ex\hbox{E}\kern-.125emX}}

\newcommand{\ie}{\textit{i.e.}}
\newcommand{\eg}{\textit{e.g.}}
\newcommand{\etal}{\textit{et~al.}}

\newif\ifComments
\Commentsfalse
\ifComments
\newcommand{\del}[1]{\noindent\textcolor{gray}{{#1}}}
\newcommand{\new}[1]{\noindent\textcolor{blue}{{#1}}}

\newcommand{\cesar}[1]{\noindent\textcolor{magenta}{Cesar: {#1}}}
\newcommand{\lucas}[1]{\noindent\textcolor{violet}{Lucas: {#1}}}
\newcommand{\sandro}[1]{\noindent\textcolor{green}{Sandro: {#1}}}
\newcommand{\guido}[1]{\noindent\textcolor{teal}{Guido: {#1}}}
\else
\newcommand{\del}[1]{}
\newcommand{\new}[1]{#1}

\newcommand{\cesar}[1]{}
\newcommand{\lucas}[1]{}
\newcommand{\sandro}[1]{}
\newcommand{\guido}[1]{}
\fi
    
\begin{document}

\title{LP-GEMM: Integrating Layout Propagation into GEMM Operations}

\ifComments
\else
\author{\IEEEauthorblockN{César Guedes Carneiro, Lucas Alvarenga, Guido Araujo, Sandro Rigo}
\IEEEauthorblockA{\textit{Instituto de Computação, Universidade Estadual de Campinas}, Campinas -- SP, Brazil \\
Emails: {\small\texttt{c261031@dac.unicamp.br, lucas.silva@ic.unicamp.br, \{guido, srigo\}@unicamp.br}}}
}
\fi

\maketitle

\begin{abstract}

In Scientific Computing and modern Machine Learning (ML) workloads, sequences of dependent General Matrix Multiplications (GEMMs) often dominate execution time. While state-of-the-art BLAS libraries aggressively optimize individual GEMM calls, they remain constrained by the BLAS API, which requires each call to independently pack input matrices and restore outputs to a canonical memory layout. In sequential GEMMs, these constraints cause redundant packing and unpacking, wasting valuable computational resources.

This paper introduces LP-GEMM, a decomposition of the GEMM kernel that enables packing-layout propagation across sequential GEMM operations. This approach eliminates unnecessary data repacking while preserving full BLAS semantic correctness at the boundaries. We evaluate LP-GEMM on x86 (AVX-512) and RISC-V (RVV 1.0) architectures across MLP-like and Attention-like workloads. Our results show average speedups of 2.25x over OpenBLAS on Intel x86 for sequential GEMMs and competitive gains relative to vendor-optimized libraries such as Intel MKL.

We demonstrate the practicality of the approach beyond microbenchmarks by implementing a standalone C++ version of the Llama-3.2 inference path using exclusively BLAS-level GEMM calls. These results confirm that leveraging data layout propagation between operations can significantly boost performance.
\end{abstract}


\section{Introduction}
Physics~\cite{karniadakis_nature_2021}, chemistry~\cite{SHI_engineering_2023}, and protein modeling~\cite{alphafold3}, as well as novel Machine Learning (ML) systems~\cite{attention,llama3,bert}, are examples of applications that leverage High-Performance Computing (HPC) techniques to make such problems tractable on modern computer systems. These applications involve a large volume of data typically represented as matrices, enabling the use of linear algebra to compute the outcomes of their mathematical models. The computation in such applications is based either on a single Matrix Multiplication (MM), a sequence of dependent MM operations, or a batch of independent MM operations (batched-MM). Furthermore, operations not immediately related to MM, such as convolutions, can be correctly computed by MM operations if the data is properly reorganized (\ie, packed) before execution. 

In computing, Matrix Multiplication (MM) is generally implemented as the General Matrix Multiply (GEMM) operation found in linear algebra libraries. It generalizes the MM operation with an additional scaling parameter $\alpha$ and a scaling-accumulation parameter $\beta$. Due to its relevance, various optimization approaches targeting different aspects of computer systems have been explored. For example, \cite{gemmini,li_frontiers_2024,kim_cal_2024} offloads MM operations to specialized hardware architectures; \cite{gotoblas, flashgemm} investigates code-level optimization techniques applied to naive implementations; and there are also numerous algorithm-level investigations of MM operations~\cite{josh_soda_2025}. Furthermore, linear algebra libraries are typically vendor-dependent. Hardware vendors, such as Intel (MKL~\cite{intel_mkl}) and NVIDIA (cuBLAS~\cite{cublas}), provide closed-source libraries with routines highly optimized for their hardware, often compliant with the Basic Linear Algebra Subprograms (BLAS) Application Programming Interface (API).

The OpenBLAS~\footnote{https://github.com/OpenMathLib/OpenBLAS} library is one of the most famous open-source efforts to provide highly optimized linear algebra routines for different architectures. In particular, it inherits from the Goto~and~Van~de~Geijn~\cite{gotoblas} GEMM optimization approach, which provides a mathematical analysis that directs a sequence of tiling, packing, and loop unrolling optimization steps over the naive MM kernel to utilize the underlying architecture more effectively. However, the BLAS API makes no assumptions about subsequent operations, simply expecting the library to provide a highly optimized implementation for the current operation. Consequently, the layout of the input data in memory must be preserved in the operation's output for consistency. Nevertheless, linear algebra workloads are rarely composed of a single operation. The Multi-Layer Perceptron (MLP) and the Transformer modules of recent ML models are examples of workloads consisting of a sequence of dependent MM operations. During execution, each MM in these modules will individually execute all steps of the library's optimized GEMM kernel (\ie, tiling, packing, and loop unrolling).

\begin{figure}[t]
    \centering
    \begin{subfigure}[t]{\linewidth}
        \centering
        \includegraphics[width=\linewidth]{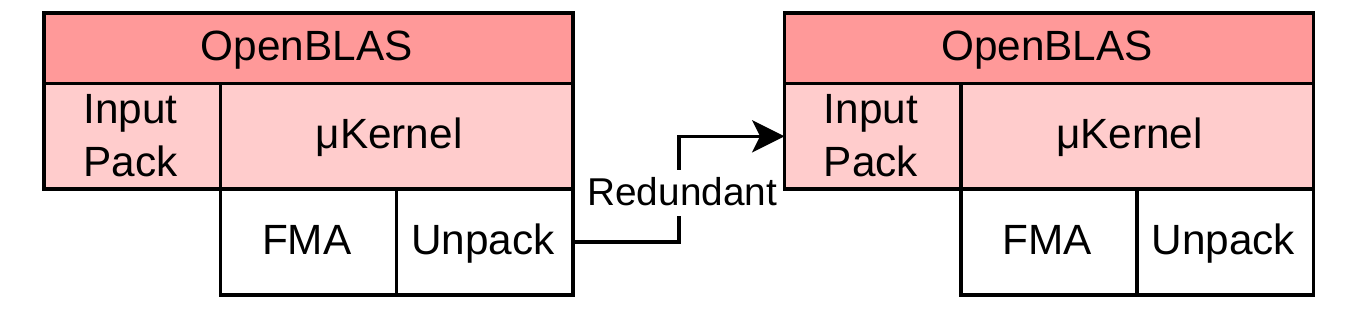} 
        \caption{Consecutive GEMM executions using OpenBLAS. The illustration depicts only two GEMMs, but the same pattern generalizes to an arbitrary number of consecutive operations.}
        \label{fig:consecutive_oblas}
    \end{subfigure}
    
    \vspace{0.6em} 
    
    \begin{subfigure}[t]{\linewidth}
        \centering
        \includegraphics[width=\linewidth]{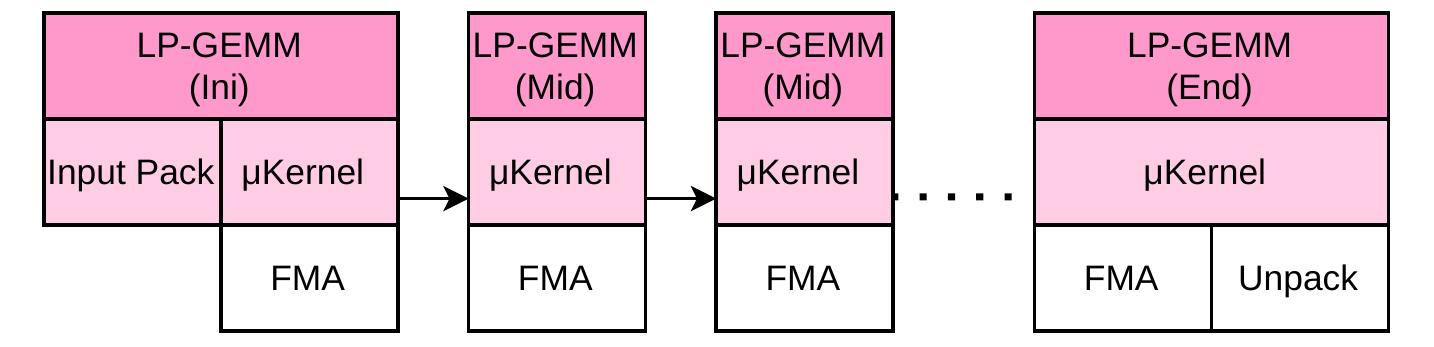}
        \caption{Consecutive GEMM executions using LP-GEMM. The figure illustrates a scenario with more than three GEMMs, where intermediate kernels (MID) are used. When only two GEMMs are executed, only the INIT and END kernels are required.}
        \label{fig:consecutive_lp}
    \end{subfigure}
    
    \caption{Comparison of OpenBLAS and LP-GEMM kernel execution of consecutive GEMM operations. The relative block sizes do not represent execution time. Weight packing is omitted for clarity.}
\end{figure}


\begin{figure*}[t]
    \centering
    \hfill
    \begin{subfigure}[c]{0.3\linewidth}
        \centering
        \includegraphics[width=.85\linewidth]{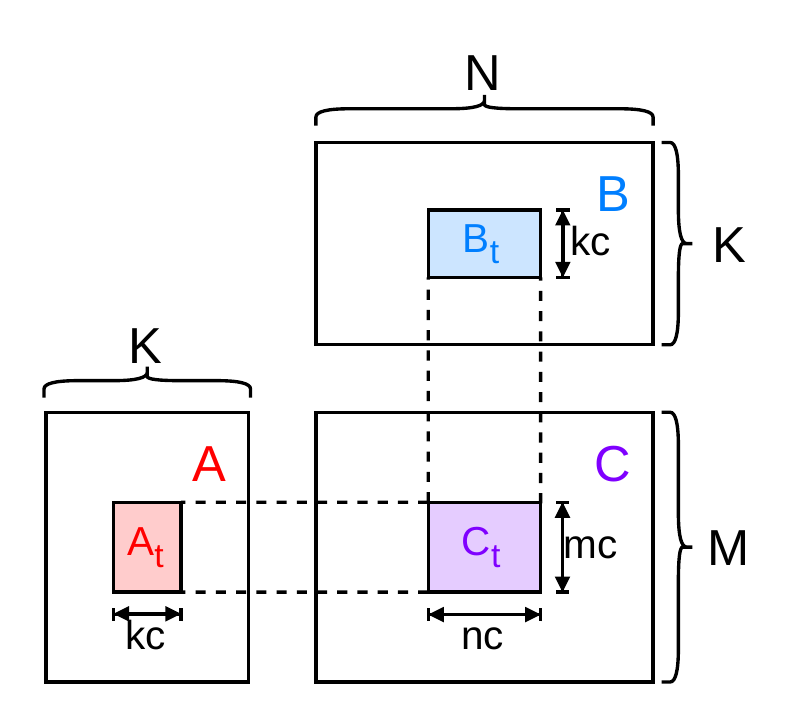}
        \caption{GotoBLAS's outer tiling.\\Kernel}
        \label{fig:gotoblasa}
    \end{subfigure}
    \begin{subfigure}[c]{0.3\linewidth}
        \centering
        \includegraphics[width=.85\linewidth]{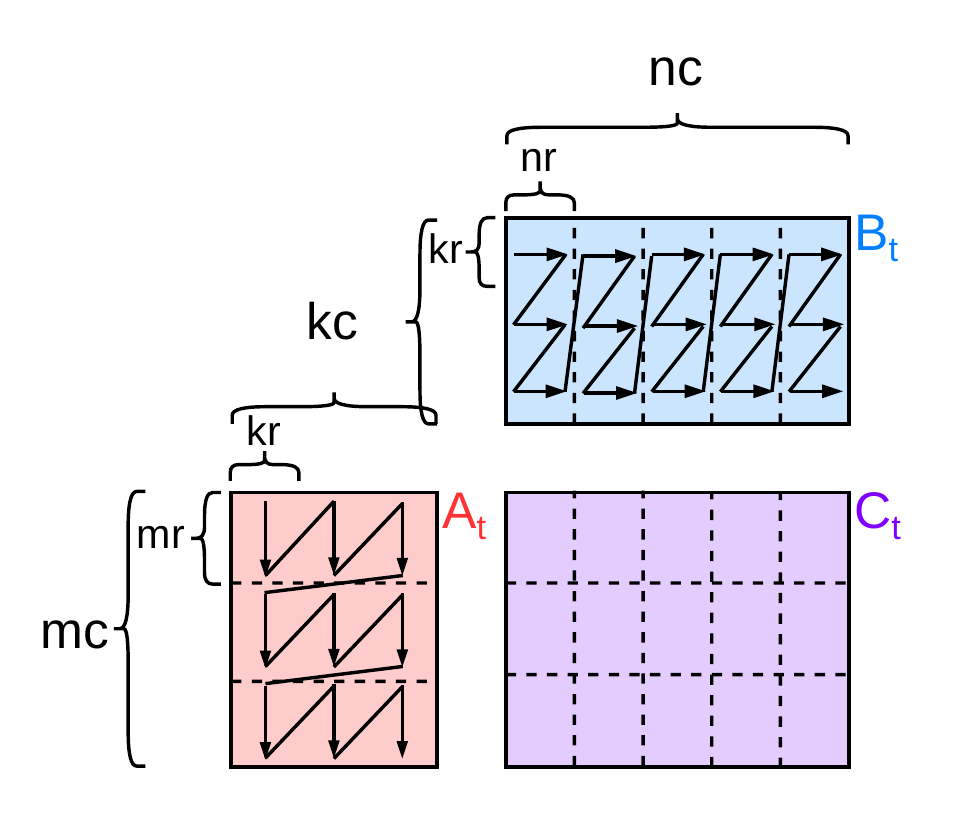}
        \caption{\parbox{.6\linewidth}{\centering \phantom{\textbf{AZ}} \newline GotoBLAS's inner tiling.\\micro-kernel (\textmu Kernel)}}
        \label{fig:gotoblasb}
    \end{subfigure}
    \begin{subfigure}[c]{0.38\linewidth}
        \hrule height 1pt 
        \medskip 

        \begin{algorithmic}[1]
            \FOR{$j = 0$ \TO $N$ \textbf{step} $n_c$}
                \FOR{$l = 0$ \TO $K$ \textbf{step} $k_c$}
                    \STATE Pack $m_c$ rows of $A_T$ into \textbf{sa}
                    \FOR{$jj = j$ \TO $j + n_c$ \textbf{step} $n_r$}
                        \STATE Pack $n_r$ columns of $B_T$ into \textbf{sb}
                        \STATE $C_T \leftarrow A_T \cdot B_T$ \COMMENT{\textmu Kernel w/ \textbf{sa}, \textbf{sb}}
                    \ENDFOR
                    \FOR{$i = m_c$ \TO $M$ \textbf{step} $m_c$}
                        \STATE Pack next $A_T$ into buffer \textbf{sa}
                        \STATE $C_T \leftarrow A_T \cdot B_T$ \COMMENT{\textmu Kernel w/ \textbf{sa}, \textbf{sb}}
                    \ENDFOR
                \ENDFOR
            \ENDFOR
        \end{algorithmic}
    
        \medskip
        \hrule height 1pt

        \caption{GotoBLAS's algorithm.}
        \label{alg:gotoblas}
    \end{subfigure}
    \hfill
    \caption{Visual representation of GotoBLAS approach~\cite{gotoblas}. (a) shows how matrices are tiled for the architecture's memory hierarchy, while (b) depicts how each block $A_t$ and $B_t$ is further tiled and packed for better usage of the processor's registers during $C_T$ computing of \textmu Kernel. Finally, (c) shows the final GotoBLAS GEMM algorithm.} \label{fig:gotoblas}
\end{figure*}

In particular, this work observes that when MM operations are executed sequentially, redundant packing steps are performed for each of them, as shown in Figure~\ref{fig:consecutive_oblas}. Therefore, we propose the Layout Propagation GEMM (LP-GEMM), a decomposition of the OpenBLAS GEMM kernel into three specialized kernels that avoid executing redundant packing steps. Specifically, the GEMM operations are divided into: (1) the Initial GEMM (ini-GEMM), which performs packing but propagates the layout; (2) the intermediate GEMM (mid-GEMM), which receives the data already packed from ini-GEMM and retains this layout for the next operations; and (3) the Ending GEMM (end-GEMM), which unpacks the data into the original layout. Figure~\ref{fig:consecutive_lp} shows how LP-GEMM can simplify the BLAS GEMM kernel, removing unnecessary operations by leveraging layout propagation between operations.

LP-GEMM was evaluated on x86 (AVX-512) and RISC-V (RVV 1.0) platforms with their vector extensions enabled, resulting in average performance speedups of up to 2x for x86 and up to 5x for RISC-V for Attention-like workloads. For isolated GEMMs extracted from real-world applications, LP-GEMM achieved results comparable to those of highly specialized, vendor-distributed BLAS libraries for x86, and a considerable speedup in relation to other state-of-the-art approaches in RISC-V.

The main contributions of this work are as follows: 
\begin{itemize} 
    \item The LP-GEMM decomposition of the OpenBLAS GEMM operation into new kernel operations, named ini-GEMM, mid-GEMM, and end-GEMM, which leverage data layout propagation.
    \item A standalone implementation of LLaMA 3.2~\cite{llama3} in C++ using the BLAS API for GEMM operations\new{~\footnote{https://github.com/czariv/simple\_attention}}. 
    \item The implementation of other common matrix operations in ML workloads (\ie, Softmax, RoPE, RMSNorm) on AVX-512, using the LP-GEMM propagated data layout.
\end{itemize}

The paper is structured as follows. First, the background of this work is discussed in Section~\ref{sec:background}. Then, our LP-GEMM approach is described in Section~\ref{sec:kernel}. Section~\ref{sec:simple_attention} describes how layout propagation was integrated into the Attention layer. LP-GEMM experimental results are presented and discussed in Section~\ref{sec:experiments}. A review of the literature is given in Section~\ref{sec:related}. Finally, concluding remarks are provided in Section~\ref{sec:conclusion}.

\section{Background}\label{sec:background}

This section provides the necessary background to understand the proposed LP-GEMM approach.

\subsection{GEneral Matrix Multiplication -- GEMM}

GEMM, as described in Equation~\ref{eq:gemm}, refers to a standard matrix multiplication (MM) of two operands $A \in \mathbb{R}^{m\times k}$ and $B \in \mathbb{R}^{k\times n}$ that produces a matrix $C \in \mathbb{R}^{m\times n}$, extended with two additional scaling parameters $\alpha$ and $\beta$. 
\begin{equation} \label{eq:gemm}
    C = \alpha A \times B + \beta C
\end{equation}

When translating a GEMM computation into code, a naive implementation may use a three-level loop nest that iterates over the $m$, $n$, and $k$ dimensions (Algorithm~\ref{alg:naive_gemm}). However, such an implementation is not optimal. It fails to effectively exploit the available memory hierarchy, causing the operands to be loaded multiple times with little to no data reuse. For example, if matrix $B$ is stored in row-major format, almost every access to its elements can result in a cache miss.

\begin{algorithm}[t]
    \caption{Naive GEMM Implementation}
    \label{alg:naive_gemm}
    \begin{algorithmic}[1]
        \REQUIRE Matrices $A \in \mathcal{R}^{m \times k}$, $B \in \mathcal{R}^{k \times n}$, $C \in \mathcal{R}^{m \times n}$, scalars $\alpha, \beta$
        \ENSURE $C \leftarrow \alpha AB + \beta C$
        
        \FOR{$i = 0$ \TO $m-1$}
            \FOR{$j = 0$ \TO $n-1$}
                \STATE $C_{i,j} \leftarrow \beta \cdot C_{i,j}$
                \FOR{$l = 0$ \TO $k-1$}
                    \STATE $C_{i,j} \leftarrow C_{i,j} + \alpha \cdot A_{i,l} \cdot B_{l,j}$
                \ENDFOR
            \ENDFOR
        \ENDFOR
    \end{algorithmic}
\end{algorithm}






There are multiple ways to improve the naive implementation. When possible, computation may be offloaded to specialized hardware~\cite{kim_cal_2024,remke_scw_2024} or optimized using HPC techniques to enhance single-core data reuse and multi-core parallelism. Goto and Van de Geijn~\cite{gotoblas} revolutionized matrix multiplication optimization with GotoBLAS by systematically analyzing GEMM’s computational intensity and its interaction with the memory hierarchy, and by organizing the computation around hardware-driven design principles. A key contribution of this approach is the systematic adoption of an \emph{outer-product} formulation as the organizing principle for register-level computation, where GEMM is expressed as a sequence of rank-1 updates that accumulate the product of a column of $A$ and a row of $B$ into tiles of $C$, enabling partial results to remain resident in registers across multiple updates.

This outer-product organization aligns naturally with modern processors that provide hardware support for Fused Multiply-Add (FMA) instructions. By expressing the update $C \leftarrow C + A_{:,l} \times B_{l,:}$ using FMA operations on small register-resident tiles, GotoBLAS maximizes arithmetic throughput while minimizing instruction overhead, while also exposing opportunities for register blocking and vectorization. These principles form the foundation of the highly optimized computation structures employed in GotoBLAS and later adopted by modern BLAS implementations. For simplicity, we assume $\alpha = 1$ and $\beta = 0$ for the remainder of the paper.

\paragraph{\textbf{Tiling}} Loop tiling, or blocking, is used to optimize nested loops. It enhances data locality within the memory hierarchy by adjusting the iteration space to operate on smaller blocks rather than the entire data set at once. To accomplish this, new outermost loop levels are created to iterate over blocks, and the iteration space of the original element-level loops is adjusted to iterate over the elements within each block~\cite{lam1991cache,wolfe_book}.

\paragraph{\textbf{Packing}} Packing optimizes memory hierarchy usage by buffering tiled blocks into contiguous memory regions and reordering their elements to match the execution stride. This minimizes Translation Lookaside Buffer (TLB) misses and reduces cache conflict misses~\cite{gpat}.

\paragraph{\textbf{Loop Unrolling}} Loop unrolling replicates the body of a loop multiple times within a single iteration. This reduces loop control overhead and exposes more independent instructions to the processor, thereby improving instruction-level parallelism and instruction scheduling~\cite{patterson_book}.

Fig.~\ref{fig:gotoblas} shows the GotoBLAS GEMM optimization strategy. The process starts with the outer tiling of the GEMM operation, referred to as the \textbf{kernel} (Fig.~\ref{fig:gotoblasa}), which tiles all MM-related dimensions using parameters $n_c$, $m_c$, and $k_c$ derived from the processor’s memory hierarchy, and parameters $n_r$, $m_r$, and $k_r$ derived from register sizes. Subsequently, as shown in Figure~\ref{fig:gotoblasb}, a \textbf{micro-kernel} (\textmu Kernel) computes a given $C_T$ tile from blocks $A_T$ and $B_T$. During computation, the $A_T$ and $B_T$ tiles are first packed (as indicated by the arrows in Figure~\ref{fig:gotoblasb}) into contiguous temporary buffers, thereby placing them at the appropriate level of the memory hierarchy. Finally, the $C_T$ results are generated according to this packed layout, requiring an unpacking step before they are written back to memory. This approach is also detailed in the algorithm in Figure~\ref{alg:gotoblas}.

\subsection{Basic Linear Algebra Subprograms -- BLAS}

The Basic Linear Algebra Subprograms (BLAS) define a standard set of routines that serve as building blocks for performing vector and matrix operations. These routines are categorized into three levels: \begin{itemize} 
    \item \textbf{Level 1:} Performs scalar, vector, and vector-vector operations. 
    \item \textbf{Level 2:} Performs matrix-vector operations. 
    \item \textbf{Level 3:} Performs matrix-matrix operations. 
\end{itemize} 
Due to their efficiency, portability, and widespread availability, BLAS routines are commonly used as the foundation for high-quality linear algebra software~\cite{netlib_blas}.

In general, different processor vendors provide linear algebra libraries that are highly tuned for their hardware, such as MKL (closed-source)~\cite{intel_mkl} and oneDNN~\cite{onednn} for Intel processors, and cuBLAS for NVIDIA devices~\cite{cublas}. An open-source alternative BLAS implementation is the OpenBLAS library. It extends the GotoBLAS optimization approach to all BLAS API operations, supporting a wide range of architectures including LoongArch, MIPS, ARM, x86, and RISC-V. It provides a common memory-hierarchy tiling strategy used for all architectures. However, the innermost tiling is guided by the processor's register sizes (\ie, the micro-kernel or $\mu$K), which is implemented by experts for each target architecture.

%
%
This work focuses on the GEMM implementation in OpenBLAS targeting Intel x86 and RISC-V processors, leveraging the AVX-512 and RVV v1.0 vector extensions, respectively. An OpenBLAS GEMM consists of input packing, computation via the \textmu Kernel, and output unpacking to restore the standard memory layout. When executing two or more consecutive matrix multiplications using OpenBLAS, where the output of one GEMM serves as the input to the next, each intermediate result is first unpacked by the producer and then immediately repacked by the consumer. As illustrated in Fig.~\ref{fig:consecutive_oblas}, this results in redundant packing and unpacking steps at every GEMM boundary, except for the final operation.

In contrast, LP-GEMM is explicitly designed to eliminate this inefficiency by enabling layout propagation across consecutive GEMM invocations. As depicted in Fig.~\ref{fig:consecutive_lp}, the Ini-kernel performs the initial packing and establishes the internal layout, which is preserved across all Mid-kernels without intermediate unpacking. Only in the End-kernel is the accumulated result converted back to the standard layout. This design minimizes unnecessary data movement and improves efficiency in workloads characterized by chains of dependent GEMM operations.




\begin{algorithm}[!b]
    \caption{Multi-head Attention}\label{alg:multi_head}
    \begin{algorithmic}[1]
        \REQUIRE Matrices $X \in \mathbb{R}^{m \times k}$, $W_q \in \mathbb{R}^{k \times k}$, $W_k \in \mathbb{R}^{k \times m}$, $W_v \in \mathbb{R}^{k \times m}$, $W_o \in \mathbb{R}^{k \times k}$, scalar $heads$
        \ENSURE Matrix $Y \in \mathbb{R}^{m \times k}$
        \STATE $Q \leftarrow X \cdot W_q$ \COMMENT{$Q \in \mathbb{R}^{m \times k}$}
        \STATE $K \leftarrow X \cdot W_k$ \COMMENT{$K \in \mathbb{R}^{k \times k}$}
        \STATE $V \leftarrow X \cdot W_v$ \COMMENT{$V \in \mathbb{R}^{k \times k}$}
        \STATE Apply rotatory position embedding (\textbf{RoPE}) to $Q$ and $K$
        \STATE Replicate $K$ and $V$ head-wise by $m/k$
        \FORALL{$head_i$ in $heads$}
            \STATE $Q_h, K_h, V_h, Y_h \leftarrow \mathbf{head}(h,Q,K,V,Y)$
            \STATE $tmp \leftarrow Q_h \cdot K_h$
            \STATE $tmp \leftarrow \mathbf{scale}(tmp) * mask$ \COMMENT{$mask$ is optional}
            \STATE $tmp \leftarrow \mathbf{softmax}(tmp)$
            \STATE $Y_h \leftarrow tmp \cdot V_h$
            \ENDFOR
        \STATE $Y \leftarrow Y \cdot W_o $
    \end{algorithmic}
\end{algorithm}

\subsection{Machine Learning Workloads}

ML workloads are dominated by a small set of highly structured key operations that typically reduce to tensor/matrix multiplications or element-wise operations~\cite{thangamani_2025_arxiv}. Liu~\etal~\cite{neocpu} categorized ML operations into three groups, depending on how they interact with the data layout: (1) operations that are oblivious to the data layout (\eg, element-wise operations); (2) operations that are layout-tolerant (\eg, GEMM and Softmax); and (3) operations that modify the data layout (\eg, Flatten and Reshape).

However, such operations do not appear in isolation. Deep ML models are composed of sets of operations with complex dependencies. A Multi-Layer Perceptron (MLP), one of the most common constructs in ML, consists of a sequence of matrix-multiplication (MM) layers interleaved with activation functions. The same holds for recent Transformer architectures~\cite{attention}, which contain projection layers, multi-head attention layers, and MLPs.

In this work, we focus on optimizing sequences of GEMMs. That is, given an input matrix $X$ and a set of $S$ weight matrices $W_i$ and activation functions $\mathcal{F}_i$, the output matrix $O$ can be computed as
\begin{equation}\label{eq:sequential_gemm}
O = \mathcal{F}_S\bigl(W_S \times \mathcal{F}_{S-1}(\cdots \times \mathcal{F}_1(W_1 \times X) \cdots )\bigr).
\end{equation}

Using this notation, we describe two primary deep learning architectures as groups of sequential GEMMs:

\begin{enumerate}
    \item \textbf{Multi-Layer Perceptrons (MLPs):} These networks follow the formulation in Eq.~\ref{eq:sequential_gemm}, where the output of one layer serves as the input to the next layer's GEMM operation.
    \item \textbf{Transformers:} The Transformer's main building block, the multi-head attention mechanism (Algorithm~\ref{alg:multi_head}), can be described as follows: (1) three initial MM operations that serve as input projections for $Q$, $K$, and $V$ (Lines 1--3); (2) the attention logic (Lines 4--9), which involves two additional MM operations, namely the score calculation ($Q \cdot K$) and the weighted summation ($tmp \cdot V$), interleaved with Softmax normalization; and (3) the final output projection (Line 10).
\end{enumerate}


\section{Propagating Layout in GEMM Operations}\label{sec:kernel}

Typical BLAS libraries are designed to operate as isolated kernels, producing outputs in the same layout as the naive implementation, leading to some redundant operations. For this reason, LP-GEMM removes those redundancies, allowing the packing layouts constructed in earlier kernels to be propagated to subsequent uses of the same matrix.

\begin{figure}[!b]
    \centering
    \includegraphics[width=0.8\linewidth]{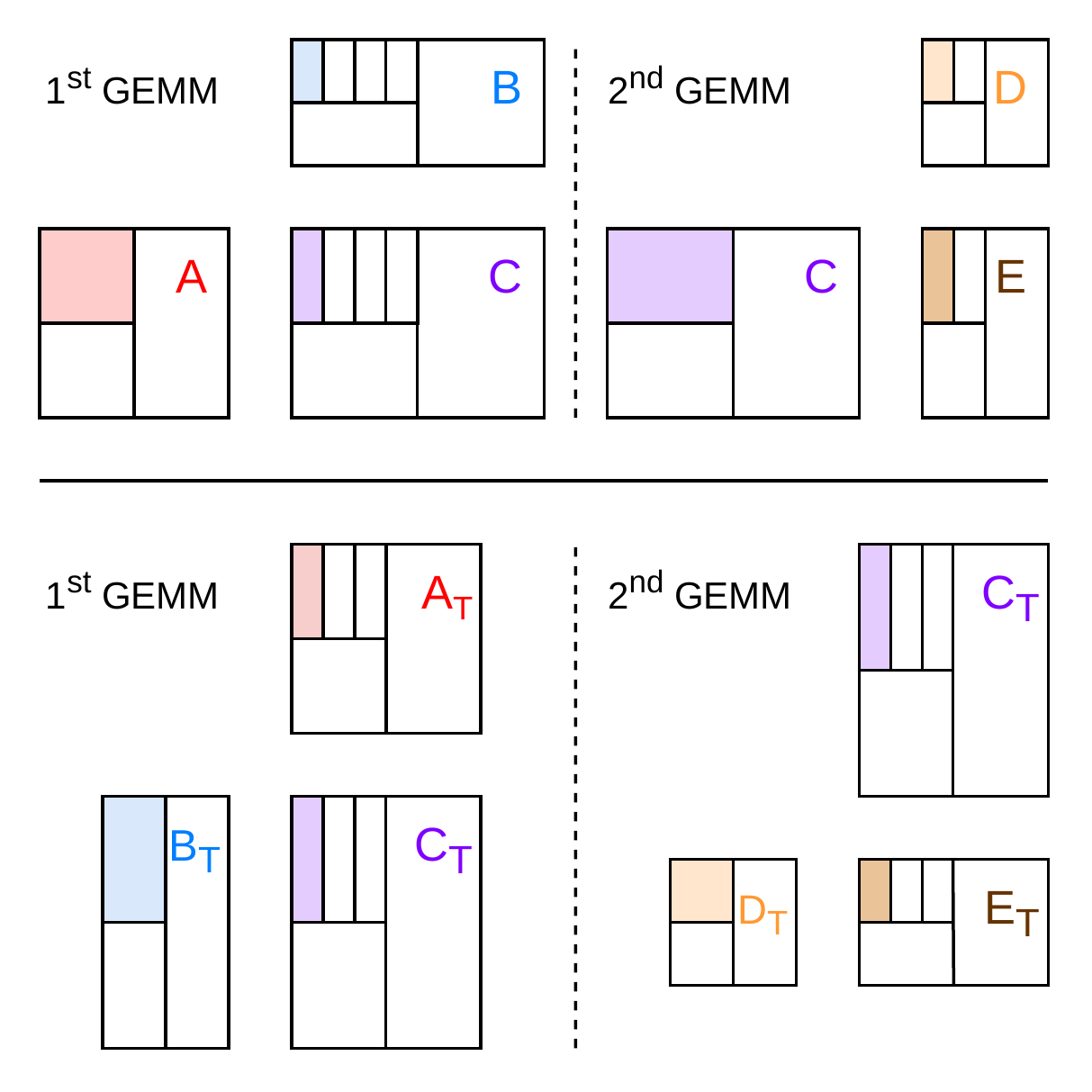}
    \caption{Overview of sequential GEMM operations using different data layouts. The results are the same as using OpenBLAS with column-major or row-major, respectively.}
    \label{fig:gemmlayouts}
\end{figure}

Figure \ref{fig:gemmlayouts} shows two ways that sequential matrix multiplications can be performed by the OpenBLAS library to generate $E$, defined by $E = D \cdot C$ with $C = A \cdot B$. This figure also illustrates the overall layout used in those distinct ways to express the sequential GEMMs. The top approach corresponds to the previous equations, where the output of one MatMul is the multiplicand of the subsequent one. The bottom approach is slightly different: by using the transposes of the matrices, it is possible to use the output of one MatMul as the multiplier of the next one. This latter approach exhibits a clear common layout between the matrix generated in the first GEMM and how it is consumed by the second GEMM, as shown by the highlighted tiles in both instances of $C_t$.

\begin{figure*}[!htp]
    \centering
    \begin{tabular}{@{}c@{\qquad}c@{}} 
    \begin{subfigure}[t]{0.4\textwidth}
        \centering
        \includegraphics[width=0.8\linewidth]{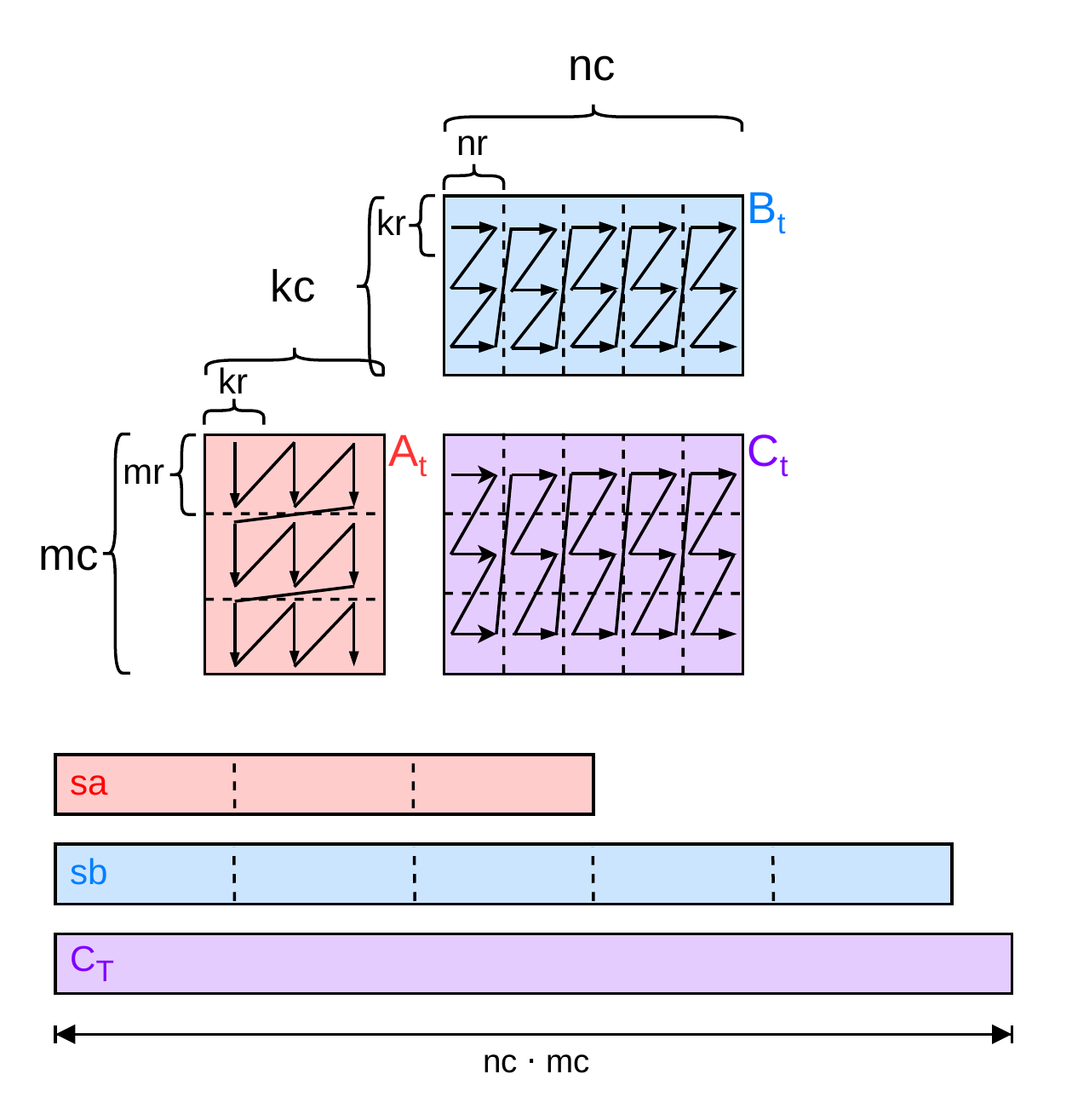}
        \caption{Micro-kernel compute order.}
        \label{fig:compute-order}
    \end{subfigure}
    &
    \begin{subfigure}[t]{0.4\textwidth}
        \centering
        \includegraphics[width=0.9\linewidth]{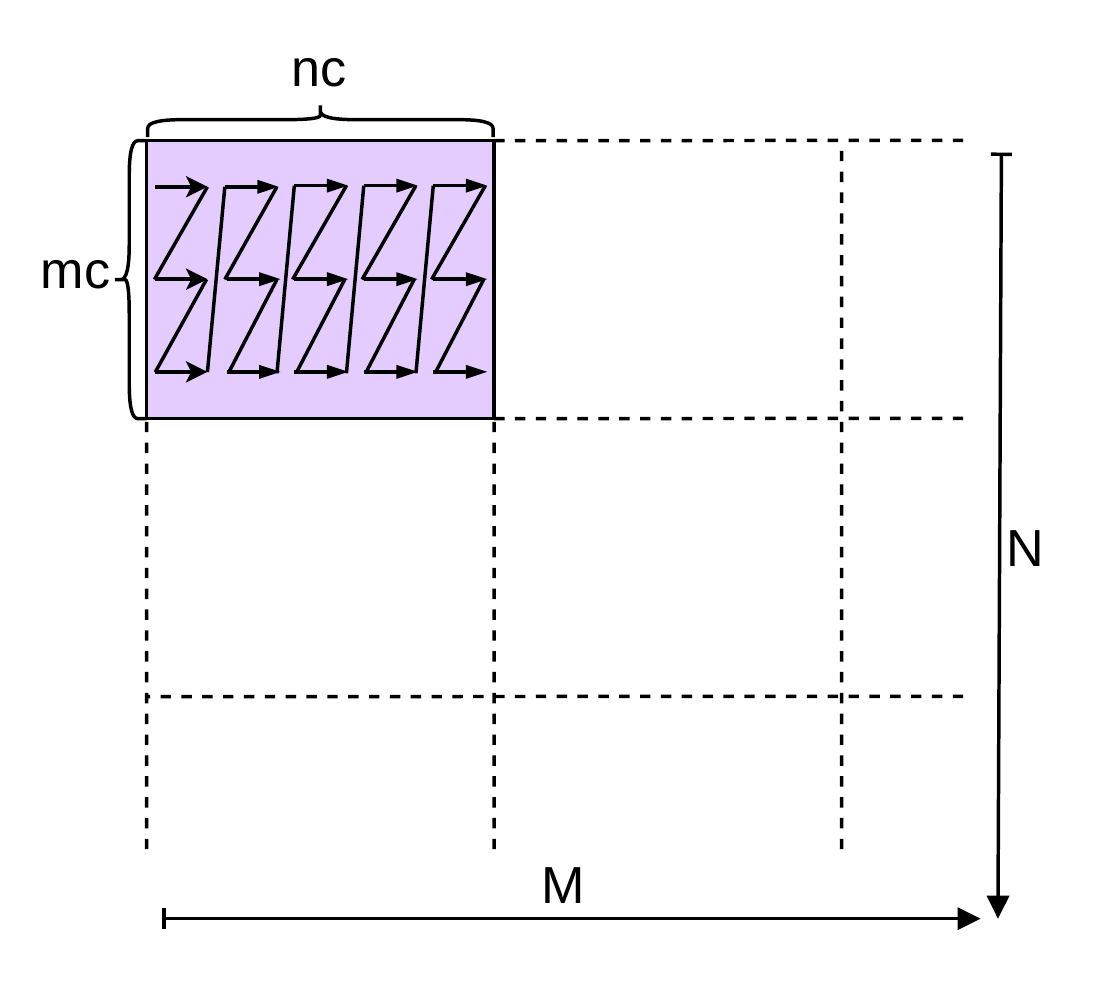}
        \caption{Default saving order, going back to original layout.}
        \label{fig:default-layout}
    \end{subfigure}
    \end{tabular}
    
    \vspace{1em}
    \begin{subfigure}{\textwidth}
        \includegraphics[width=\linewidth]{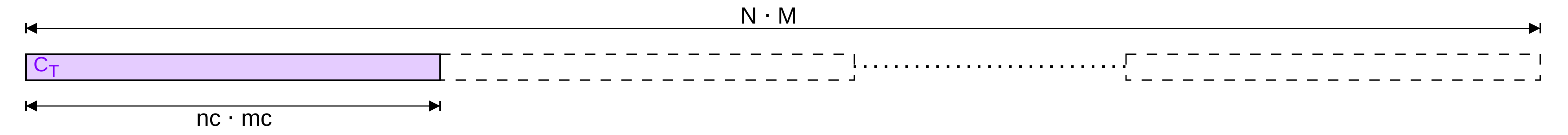}
        \caption{Contiguous storage of C, propagating packing layout.}
        \label{fig:prop-layout}
    \end{subfigure}
    \caption{Micro-kernel Layout.}
    \label{fig:micro-kernel-layout}
\end{figure*}

By design in OpenBLAS, the output shares a similar tile structure to the one used by the multiplier, as illustrated by the arrow showing $C_T$ compute order in Fig.~\ref{fig:compute-order}. This structure allows for a very straightforward layout propagation, which LP-GEMM can leverage. Alternatively, if matrix $E$ is produced using the traditional approach to sequential MatMuls, storing the output matrix would require complex packing, generating additional overhead. For this reason, this paper will assume the use of transposed matrices.

\subsection{Kernel Implementation}

All kernels used by LP-GEMM are derived from the OpenBLAS kernels, which we refer to as the default implementation for both the kernel and the micro-kernel. For reference, Figure~\ref{alg:gotoblas} provides a pseudo-code representation of how OpenBLAS performs matrix multiplication.

To understand why layout propagation is not supported in BLAS-style GEMMs, it is important to distinguish three concepts:

\begin{enumerate}

    \item Packed layout used for reading inputs, which is determined by the packing strategy and specifies the layout expected by the micro-kernel.
    \item Iteration order used during computation, which dictates the order in which sub-blocks of the output are produced.
    \item Final stored layout, i.e., the layout in which the output matrix is written to memory and returned to the user.
\end{enumerate}

BLAS does not mandate a specific computation order for GEMM (i.e., it does not constrain (2)), and it also does not require any particular packed layout (1), since packing is an internal optimization detail. BLAS only prescribes the final visible output layout (3).

However, in most BLAS implementations based on GotoBLAS, (1) and (2) are tightly coupled: the packed layout (1) determines how the micro-kernel accumulates partial results, which, in turn, defines the computation order (2). Meanwhile, the final stored layout (3) must conform to the BLAS API and, therefore, often differs from both (1) and (2). As a consequence, implementations must perform an additional unpacking/reordering step to convert the output produced according to (2) into the required layout (3). This relationship is illustrated in Figures~\ref{fig:compute-order} and~\ref{fig:default-layout}.

LP-GEMM aims to unify these three aspects -- making (1) = (2) = (3), so that no packing or unpacking-related reordering is necessary. By aligning the packed layout, the computation order, and the final stored layout, LP-GEMM enables the output of one GEMM to be consumed directly by the next without redundancy. Data can then be stored sequentially in memory using this common layout, drastically boosting spatial locality, as demonstrated in Figure~\ref{fig:prop-layout}. This design choice, however, violates the BLAS requirement that results be returned in a specific canonical layout. Therefore, LP-GEMM is not compliant with the BLAS API.

To safely propagate layouts across GEMM operations while preserving correctness, LP-GEMM defines three types of kernels: the \textbf{Initial Kernel}, the \textbf{Intermediate Kernel}, and the\textbf{ Ending Kernel}. These kernels rely on two micro-kernels: the \textbf{Propagate-Layout \textmu kernel}, which maintains the propagated layout across GEMMs, and the \textbf{Default \textmu kernel}, which terminates propagation and restores the BLAS-style output layout when needed.

The following subsections explain the modifications made to the original kernel, detailed in Figure~\ref{alg:gotoblas}, in order to create three different kernel versions that together implement the layout propagation strategy.

\subsubsection{Initial Kernel}
This kernel should come before all the others, as it is responsible for starting the layout propagation. The kernel itself is not much different from the one in OpenBLAS. The only difference is that it uses the propagate layout micro-kernel to perform the GEMM, which leads to the output being stored in the propagated layout of LP-GEMM.

\subsubsection{Intermediate Kernel}
This kernel assumes that the multiplier is already packed into the layout that will be used by the micro-kernel, allowing the kernel to skip packing this matrix. For this reason, this kernel can only be called after packing the multiplier, which can be done either by an Initial Kernel or by directly packing it before calling this kernel. Besides that, this kernel continues the layout propagation by storing its output in the same order it is calculated, which allows for most of the GEMMs in a sequence of MatMuls to be solved by this kernel. The difference in implementation comes from removing lines 3-7 and starting the loop of line 8 at position 0 instead of $mc$.

\subsubsection{Ending Kernel}
This kernel is responsible for stopping the layout propagation by producing an output in the original data layout. This kernel is structured the same way as the Intermediate Kernel, with the same changes to the implementation. The difference is that it uses the Default \textmu kernel, which itself resets the data back to the original layout.

\subsection{Micro-kernel Design}

Both the Propagate Layout and the Default \textmu kernels use the same structure to calculate the values of $C$; the arrow in Figure~\ref{fig:compute-order} demonstrates the order in which the values of $C$ are calculated. The default \textmu kernel implementation, shown in Figure~\ref{fig:default-layout}, has to save $C$ in an order different from the one in which it is computed, so that the resulting output has the same layout as the original operator.

The LP-GEMM layout propagator micro-kernel takes advantage of the notion that $C$ is computed in a \textit{packed} layout, which can be expressed as:
\begin{equation}\label{eq:prop-layout}
    \frac{N}{nc}\frac{M}{mc}\frac{nc}{nr}\frac{mc}{mr}\;nr\;mr
\end{equation}

This means that data is contiguous in the $nr\times mr$ dimension, which is important for micro-architecture-specific register population. But most importantly, $C$ will be contiguous in the $nc\times mc$ dimension, which is the same dimension that will be used to calculate data in the case of a subsequent consumer GEMM operator.

As data is not stored in its original layout, it is required that the following are true: (1) If $C$ already contained data before the first GEMM call, and $\beta$ is not $zero$; $C$ must be packed into the layout of equation~\ref{eq:prop-layout}. (2) Calls to the \textmu kernel need to take into account that $C$ will not be stored in the original layout, and the regions of this matrix passed to the micro-kernel need to reflect this notion. The LP-GEMM kernels that use layout propagation (e.g., initial and intermediate) already solve (2).

The scenario described by (1) cannot be solved by LP-GEMM without significant changes to the kernel and \textmu kernel implementations. Luckily, scenarios like this are rare, especially in AI models, which are the primary motivators for this work. For this reason, we decided not to address this corner case in our proposed implementation.

\subsection{Strided Loads and Stores}\label{sec:stride}
The flexibility of matrix multiplications means that the use cases for it are numerous, and LP-GEMM had to be able to replicate naive GEMM behavior for those cases. Although one of the most common ways to think of an MM is to consider multiplying contiguous matrices in memory, it is not always the case, as sparse or tiled matrices may be involved.

The OpenBLAS kernel can seemingly deal with these cases through the manipulation of kernel parameters such as M, N, K, and the sizes (lda, ldb, and ldc) of the matrices. This is possible because, by default, data is stored requiring some stride in memory accesses in those kernels.

Because LP-GEMM stores data in the same order in which it is computed, sometimes such behavior cannot seemingly be adapted to work in those kernels. In response to this, it is necessary to use some additional function parameters for those scenarios. One of those parameters is simply responsible for changing the order of the blocks $C_T$ that will be saved sequentially. The second defines a stride for data to be stored in memory, so that a GEMM call leaves some empty positions between data from a single block $C_T$, crucial when calculating only a fraction of the output matrix.

\section{Layout changes in the Attention Layer}\label{sec:simple_attention}
To test the generalizability of LP-GEMM, we developed a mock of the attention layer of the LLaMA transformer using C++, so that the layout could be propagated freely while testing the correctness of the results compared to the baseline execution of the actual model. To assess the validity of this implementation, the randomly generated inputs, all intermediate calculations, and the weights of the LLaMA 3.2 (1B parameters)~\footnote{https://huggingface.co/meta-llama/Llama-3.2-1B} version were exported from the PyTorch implementation to binary files. These same inputs and weights were read by the C++ implementation, and all intermediate results were compared to those extracted from PyTorch.

\subsection{Layout Propagation in Matrix Operations}\label{sec:lp_matrix}
A key feature of LP-GEMM is its ability to propagate the packing layout from one GEMM operation to the next. In practice, however, back-to-back matrix multiplications are uncommon. Machine-learning workloads illustrate this well, as after almost every matrix multiplication, an activation or other non-linear transformation is applied. To evaluate how LP-GEMM interacts with such operators, we use the attention layer of transformer models as a representative scenario.

Multi-head attention performs several matrix multiplications that can be grouped as a sequence of GEMMs, making it a suitable candidate for LP-GEMM. Layout propagation across these operations, however, is subtle. Because the $Q$, $K$, and $V$ matrices are split into multiple heads, each head consumes different slices of the original matrices. This is inconsequential without layout propagation (see Section~\ref{sec:stride})\lucas{?}, but LP-GEMM stores its output in the exact layout expected by the consumer GEMM. As a result, the kernel must know the correct strides to generate a layout compatible with downstream consumers. Fortunately, these strides are highly predictable, depending only on the input dimensions and the per-head projection size, allowing the compiler to infer them statically.

As shown in Algorithm~\ref{alg:multi_head}, several matrix operations appear between the GEMMs in the attention layer. These include: Scale, which is layout-oblivious; and RoPE and Softmax, which are layout-tolerant but require adjustments to preserve correctness under a propagated layout. The problem is that the optimal layout for these operators differs from the one preferred by the subsequent consumer GEMMs. A naive address translation would introduce substantial overhead due to irregular memory access. It is therefore necessary to adapt these intermediate operations so that they better exploit the LP-GEMM layout.

\paragraph{Rotatory Position Embedding (RoPE)} naturally aligns with the layout produced for the following GEMM. Rotations are applied independently per row within each head, which is the same pattern used when the LP-GEMM kernel loads data. As a result, RoPE can operate efficiently on the propagated layout with minimal modification. However, it can actively produce better results if multiple rows are calculated simultaneously using SIMD, taking advantage of the row interleaving done in the propagation layout.

\paragraph{Softmax} poses a greater challenge. It is computed independently for each row, whereas the propagated layout partitions the row dimension into tiles for outer-product processing. Direct propagation, therefore, introduces address-jump overhead at tile boundaries. However, because the propagated layout interleaves rows, Softmax can be reorganized to operate over multiple rows at once, reducing the penalty associated with non-sequential accesses.

More generally, other matrix operations can also be adapted to operate directly on the LP-GEMM layout, and the compiler can determine the required transformations at compile time. For row-major operations that rely on full-row reduction and cannot be decomposed into partial results, such as Softmax, some overhead is expected when using the propagated layout of Equation~\ref{eq:prop-layout}. Nonetheless, as demonstrated in Section~\ref{sec:attn_layer}, the performance gains from LP-GEMM outweigh these additional costs.

\section{Evaluation of LP-GEMM}\label{sec:experiments}

This section evaluates LP-GEMM along two primary dimensions: performance and generalizability. First, the experimental setup is described, including the target x86-64 and RISC-V platforms and the set of optimized GEMM implementations used for comparison. Next, we assess performance using isolated GEMM executions, comparing LP-GEMM against state-of-the-art approaches on both architectures. We then evaluate LP-GEMM within the Attention layer of the LLaMA 3.2 model to study its behavior in a realistic machine learning workload. Finally, we analyze a sequence of three consecutive GEMM operations to compare LP-GEMM with existing approaches for optimizing sequences of matrix multiplications.

\begin{table}[t]
    \begin{tabular}{lll}
        \hline
        CPU info                      & Intel Xeon Gold 6252 & Spacemit X60 \\ \hline
        CPU Family                    & Skylakex (x86-64)    & RV64GCVB     \\
        Vector Extension              & AVX512               & RVV1.0       \\
        $m_c, n_c$ and $k_c$          & 448, 16384, 448      & 128, 16385, 128 \\
        $m_r$ and $n_r$               & 16, 4                & 16, 8    \\
        L1 cache size (KiB)           & 32                   & 32           \\
        L2 cache size (KiB)           & 1024                 & 512 shared   \\
        L3 cache size (KiB)           & 35.8 shared          & -            \\
        CPU Frequency           & 3.6                  & 1.6            \\
        Vector Registers Size         & 512                  & 256          \\
        FMA Throughput (GFLOPS/s)         & 223.3                  & 25.6          \\ \hline
    \end{tabular}
    \caption{Summary of evaluated systems.}
    \label{tab:execution_env}
\end{table}

\subsection{Experimental Setup}

The experiments aim to evaluate LP-GEMM on two representative architectures that provide explicit vector register support and enable Single Instruction Multiple Data (SIMD) execution, which is standard for high-performance GEMM implementations. A summary of the architectural characteristics relevant to GEMM optimization is provided in Table~\ref{tab:execution_env}.

\subsubsection{Execution Environment}

\paragraph{\textbf{RISC-V}}
Experiments on the RISC-V platform were conducted using a Banana Pi BPI-F3 equipped with a SpacemiT K1 octa-core X60 processor (RV64GCVB). This processor supports SIMD execution through the RISC-V Vector Extension (RVV) version 1.0.

\paragraph{\textbf{Intel x86}}
Experiments on the x86 platform were conducted using an Intel Xeon Gold 6252 processor, which provides native SIMD support via the AVX-512 instruction set.

All performance evaluations use OpenBLAS as the primary baseline implementation. In addition to LP-GEMM, we compare against several widely used and state-of-the-art GEMM implementations, including BLIS~\cite{blis}, Intel MKL~\cite{intel_mkl}, oneDNN~\cite{onednn}, and FlashGEMM~\cite{flashgemm}. Due to platform support constraints, Intel MKL, oneDNN, and FlashGEMM are evaluated only on the x86-64 platform. LP-GEMM, OpenBLAS, and BLIS were evaluated on both x86-64 and RISC-V architectures.

All implementations are invoked through the BLAS Level-3 GEMM interface whenever it is supported. For LP-GEMM and FlashGEMM, minor adaptations were required to ensure functional equivalence during evaluation, as these implementations do not strictly conform to the standard BLAS API. All experiments were conducted in a single-threaded configuration to avoid introducing bias due to differences in parallel runtime behavior.

\subsection{Single GEMM Execution}\label{sec:single_execution}

\begin{figure}[!t]
    \centering
    \begin{subfigure}[t]{\linewidth}
        \centering
        \includegraphics[width=.8\linewidth]{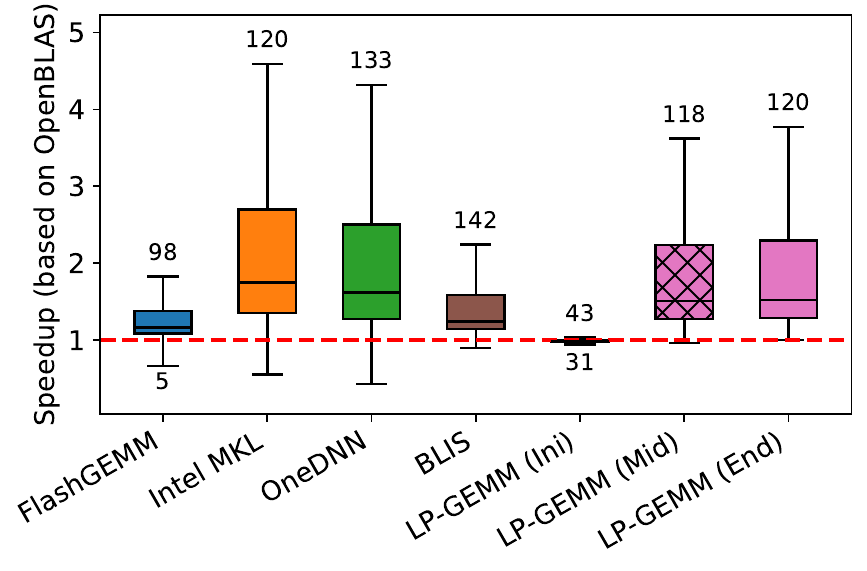}
        \caption{Intel Xeon Gold 6252}
        \label{fig:single_gemm_x86}
    \end{subfigure}

    \begin{subfigure}[t]{\linewidth}
        \centering
        \includegraphics[width=.8\linewidth]{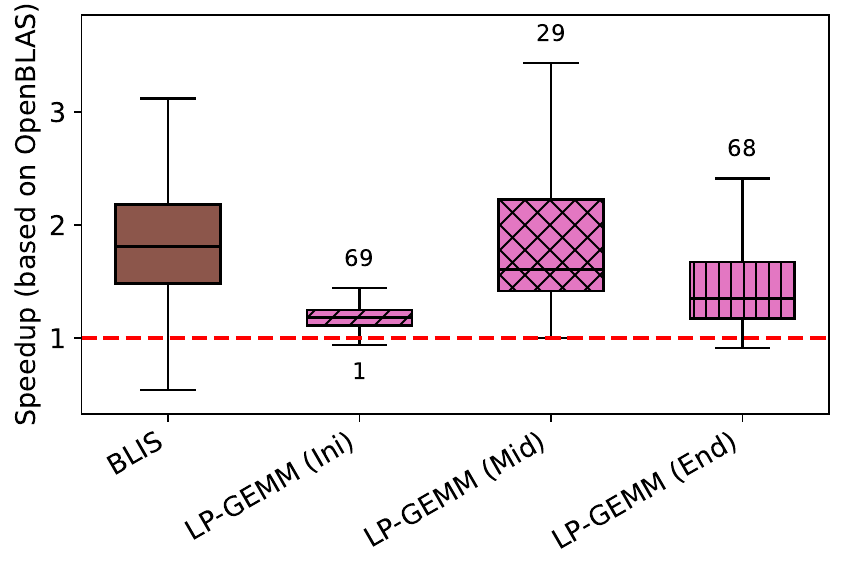}
        \caption{Spacemit X60}
    \label{fig:single_gemm_riscv}
    \end{subfigure}
    \caption{Speedup of a single GEMM extracted from gemmbench\new{~\cite{gemmbench}} using different state-of-art kernels compared to the three separate LP-GEMM kernels.}
    \label{fig:single_gemm}
\end{figure}

To evaluate the performance of the proposed kernels against state-of-the-art GEMM implementations, we conducted a single-GEMM execution experiment using matrix sizes drawn from the gemmbench~\new{~\cite{gemmbench}} dataset. Figure~\ref{fig:single_gemm} shows the results for an x86 and riscv platforms. The points used to create the boxplot represent the means of 10 executions for each implementation using each of the matrix sizes from gemmbench. All speedups are computed relative to OpenBLAS. The simplicity of this setup allows for the comparison of a wide range of implementations under the same execution model. Although LP-GEMM kernels are not designed for this usage pattern, as they expect data to already be in the propagated layout, this experiment still provides a useful perspective. It offers a lower-bound estimate of the potential gains from LP-GEMM in realistic pipelines and helps clarify where its performance advantages originate.

The Initial kernel (Ini) delivers performance close to OpenBLAS. This is expected, since, aside from removing the unpacking step and slightly altering the store pattern, it largely mirrors the baseline behavior. For large matrices, these changes alone are not sufficient to generate substantial speedups.

\begin{figure*}[!hbt]
    \centering
    \begin{subfigure}[t]{0.45\linewidth}
        \includegraphics[width=1\linewidth]{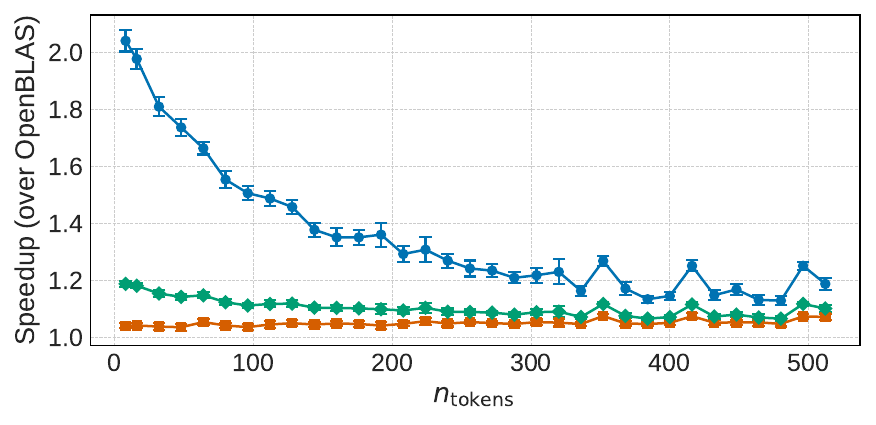}
        \caption{Intel Xeon Gold 6252}
        \label{fig:plot_attn}
    \end{subfigure}
    \hfill
    \begin{subfigure}[t]{0.45\linewidth}
        \includegraphics[width=1\linewidth]{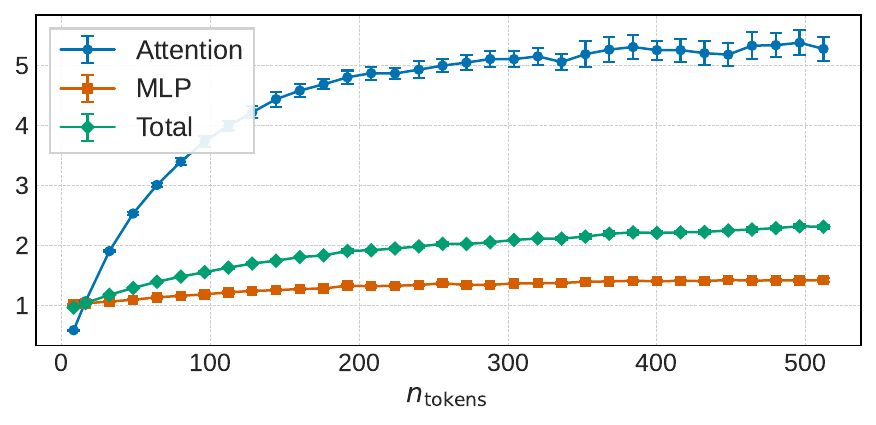}
        \caption{Spacemit X60}
        \label{fig:plot_attn_riscv}
    \end{subfigure}
    \caption{Speedup of Attention layer from LLaMA 3.2 model, with embedded dimension of 2048, and MLP weights with dimension of 8129. Comparing LP-GEMM and layout propagation to OpenBLAS with no propagation.}
    \label{fig:attn}
\end{figure*}

The Intermediate and Ending kernels (Mid and End) reveal the real benefit of layout propagation for both RISC-V and x86. By fully eliminating the packing of one of the matrices, they achieve a median speedup of about 1.5× over OpenBLAS, with third-quartiles exceeding 2×. These gains highlight that packing, rather than the micro-kernel itself, is often the dominant cost and show how impactful layout propagation can be for optimizing GEMM execution.

The LP-GEMM kernels are derived from the open-source OpenBLAS implementations, so improvements relative to OpenBLAS mainly come from removing packing rather than from deeper micro-kernel tuning. This contrasts with libraries such as BLIS~\cite{blis} and FlashGEMM~\cite{flashgemm}, whose median speedups exceed those of the Initial LP-GEMM kernel because their underlying kernels are more heavily optimized. Still, when compared to the Mid and End kernels, both BLIS and FlashGEMM show noticeably lower speedups due to their packing overhead.

Highly tuned commercial and vendor libraries like Intel MKL and OneDNN exhibit strong performance for x86, clearly outperforming OpenBLAS (as seen in Fig.~\ref{fig:single_gemm}). Their advantage over all three LP-GEMM kernels—particularly the Initial one—comes primarily from superior kernel optimization. Nevertheless, the gap between these libraries and LP-GEMM’s Mid/End kernels is modest, suggesting that layout propagation alone can close a substantial portion of the performance difference relative to highly specialized implementations.

A final advantage of LP-GEMM is its generality: layout propagation is orthogonal to micro-kernel design and can be combined with more advanced kernels to achieve similar or greater speedups. For example, if the kernels used by MKL were open-source, LP-GEMM could be integrated with them, likely yielding even stronger results.

\subsection{Attention Layer}\label{sec:attn_layer}

One of the prime motivators for this paper is layout propagation in the attention layer of a transformer network. This experiment aims to test how LP-GEMM performs in this situation and how it interacts with other matrix operations, as described in section~\ref{sec:lp_matrix}. Figure \ref{fig:plot_attn} shows the performance of the attention layer implementation proposed in section~\ref{sec:simple_attention} using LP-GEMM and modified matrix operations to deal with layout propagation, compared to using OpenBLAS with no layout propagation, executed on x86. Results in both graphs of Figure~\ref{fig:attn} were calculated as the mean of 50 analogous executions, for both architectures.

The x86 results obtained in this experiment highlight a crucial aspect of LP-GEMM and layout propagation as a whole: the improvements in layout propagation scale inversely with the size of the matrices involved in the GEMM operations. This aspect is to be expected, as packing operations have a complexity of $O(n^2)$, while the micro-kernel that performs the multiplications has a complexity of $O(n^3)$ \lucas{why ?}. This characteristic is not inherently a downside to layout propagation optimization, but it is something that needs to be considered when dealing with this type of optimization.

Taking that into consideration, the results obtained by integrating LP-GEMM into the attention layer are still impressive, with speedups of up to 2x for Multi-Headed Attention with small input sizes and still significant improvements for larger inputs. The curve for speedup in the multi-headed attention layer has an exponential-like \lucas{can you confirm it ?} shape due to some operations performed in this region using a matrix of dimensions $num\_tokens \times num\_tokens$. The curve for the MLP speedup tends to stay around 1.1, as the operations performed in this region mostly depend on the dimensions of the weights used.

The graphs focus on sequence lengths of up to 500 tokens, as, in practical Edge AI workloads, latency constraints often preclude massive context windows. Furthermore, recent empirical studies suggest that extending context length indefinitely is not strictly beneficial for reasoning tasks. Levy~\etal~\cite{same-task} demonstrated that reasoning performance on identical tasks degrades significantly as input length increases, and Du~\etal~\cite{context-length} found that this "distraction" effect persists even when retrieval is perfect. Consequently, optimizing for moderate context windows, where the hardware is most efficient and the model is cognitively most effective, remains the primary objective for edge deployment.

The RISC V results, shown in Figure~\ref{fig:plot_attn_riscv}, follow a noticeably different trend. The absolute performance gap between RISC-V and x86 is largely explained by architectural differences. The processor in the Banana Pi BPI F3 provides significantly lower FMA throughput and memory bandwidth compared to the Intel Xeon Gold 6252, which naturally limits GEMM performance due to its high arithmetic intensity. Beyond the SpacemiT K1 used in this experiment, other RISC-V implementations are targeting similar workloads, such as the Kendryte K230 (featuring a dual-core C908 with RVV 1.0 for AIoT) and the SiFive Intelligence X280, which is increasingly adopted for AI acceleration in automotive and datacenter offload scenarios.

However, the shape of the speedup curve on RISC V is governed by a different factor. The nearly linear improvement observed as $n_{tokens}$ increases is primarily a consequence of inefficiencies in the baseline RISC-V kernel used by OpenBLAS. This kernel performs the final unpacking step through out-of-order memory accesses, which become increasingly costly as matrix sizes grow. LP-GEMM avoids this overhead entirely by producing the output directly in a contiguous layout. As the penalty from scattered accesses accumulates in the reference implementation, the relative advantage of LP GEMM increases proportionally with problem size.

In summary, the hardware differences explain the performance gap between the two architectures, but the scaling behavior on RISC-V is driven mostly by the inefficiency of the reference kernel in OpenBLAS. This makes layout propagation especially effective on this platform as the size of the problem increases, in contrast to x86, where speedups tend to saturate once packing and unpacking costs become negligible relative to the highly optimized micro-kernels.

\subsection{Consecutive GEMMs}

As previously mentioned, it is uncommon to find a sequence of pure consecutive GEMMs in machine learning, as the structurally adjacent matrix multiplications (such as those found in DNN bottleneck blocks) are typically separated by essential non-linear activation layers, which break data-flow continuity. Nevertheless, other work has established benchmarks composed of this challenging sequence of three consecutive GEMM operations by extracting the input/output matrix sizes from the convolutional layers found in common DNN frameworks \cite{flashgemm}. This methodology creates a specialized, linear benchmark scenario that enables the study of fusion across multiple matrix multiplications by abstracting away the intermediate non-linear operations inherent in the original models.

\begin{figure}[htb]
    \centering
    \includegraphics[width=.9\linewidth]{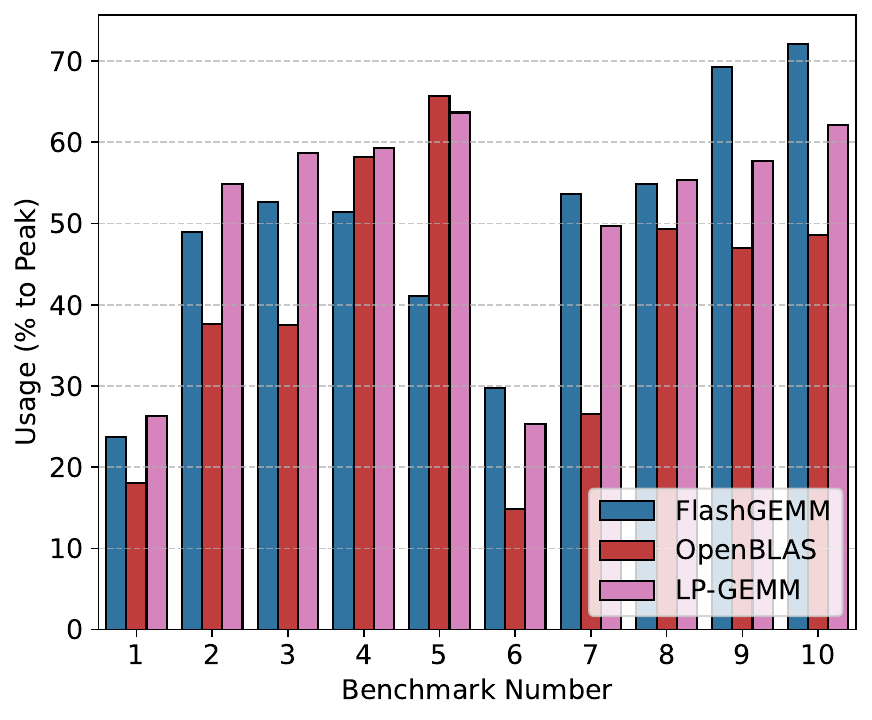}
    \caption{Usage comparison of LP-GEMM with FlashGEMM and OpenBLAS, using benchmarks extracted from DNN~\cite{flashgemm}.}
    \label{fig:flashgemm_comparison}
\end{figure}

In the benchmarks used for this experiment, most of the execution is spent on the first GEMM in the case of LP-GEMM, using the Initial Kernel. As mentioned in Section~\ref{sec:single_execution}, most of the speedup from LP-GEMM is in the other two kernels. In other words, this is not the best scenario for LP-GEMM, but we still want to compare our approach with FlashGEMM to understand whether layout propagation can bring some benefits even in these circumstances. As depicted in Fig.~\ref{fig:flashgemm_comparison}, LP-GEMM is still able to outperform both OpenBLAS and FlashGEMM on most of the proposed benchmarks. Once again, it demonstrates the power of layout propagation in a sequence of GEMM operations.

\section{Related Work}\label{sec:related}
Matrix multiplication is a fundamental building block of modern machine learning workloads and remains one of the most heavily used operations across HPC applications. Historically, GEMM optimization has focused on maximizing the performance of individual calls, treating each GEMM as an isolated computation. This design philosophy is exemplified by traditional BLAS libraries, whose implementations, often derived from GotoBLAS~\cite{gotoblas}, perform packing and unpacking on every GEMM invocation, regardless of the broader workload structure. Even highly specialized libraries such as Intel MKL, which exploit extensive low-level tuning to achieve high processor utilization, optimize each GEMM independently and do not account for opportunities across consecutive operations or propagate data layouts through a computation pipeline.

Data layout propagation itself is not new, particularly in contexts involving tensor graphs and deep-learning operators. NeoCPU~\cite{neocpu} proposes a graph-level framework that identifies profitable data layouts and determines regions where such layouts can be maintained. However, its approach is tailored specifically to ML workloads and does not generalize easily to other domains. Moreover, its layout selection relies on an analysis pass and a search procedure, creating a trade-off between search cost and layout quality that may result in suboptimal layouts being chosen.

OneDNN~\cite{onednn} provides one of the most mature implementations of layout propagation. When enabled, the framework infers optimized data formats to be used in subsequent operators. While originally optimized for Intel processors, OneDNN has since become an open-source, architecture-agnostic project. Like NeoCPU, it depends on analysis to identify propagation boundaries and requires users to specify where reordering is permitted, relying on explicit reorder primitives to transition between layouts. In contrast, LP-GEMM integrates the initial reorder step directly into the GEMM operation itself via the Initial kernel, reducing the overhead and complexity of managing layout transitions.

FlashGEMM~\cite{flashgemm}, similar in spirit to LP-GEMM, targets workloads consisting of sequential GEMMs rather than optimizing each GEMM call in isolation. Its novelty lies in an aggressive fusion of GEMMs to exploit data reuse and locality, thereby enabling layout propagation across fused operations. FlashGEMM also includes profitability analysis to determine whether packing should be performed based on memory constraints. However, its reliance on fusion makes integration into real ML workloads challenging: intermediate non-GEMM layers are not directly supported and would need to be reimplemented inside fused kernels to preserve correctness. Moreover, the fusion strategy disallows partial results, limiting the set of matrix dimensions that can be tiled efficiently.

\section{Conclusion and Future Work}\label{sec:conclusion}

This work introduced LP-GEMM, a general and effective approach for propagating data layouts across sequences of GEMM operations. By exploiting the natural layout continuity present in many workloads, LP-GEMM improves performance without requiring low-level microarchitectural tuning. Across the evaluated architectures and workflows, LP-GEMM consistently delivers speedups comparable to those of highly specialized GEMM implementations, underscoring its portability and practical relevance.

Our results highlight the importance of reducing redundant work, particularly unnecessary packing and unpacking, when optimizing matrix operations. The performance gains achieved by LP-GEMM stem entirely from eliminating these overheads, demonstrating that understanding how GEMMs interact within broader computation pipelines can unlock significant efficiency improvements beyond isolated kernel optimization.

LP-GEMM preserves a structure close to BLAS-style GEMMs to ease adoption, but it is not a drop-in replacement, as it introduces additional parameters for strided memory accesses (Section~\ref{sec:stride}) and produces outputs in a propagated layout. Nevertheless, integration into existing code requires only modest changes, as demonstrated in the Attention-layer case study (Section~\ref{sec:attn_layer}).

Although LP-GEMM is currently implemented for x86\_64 and RISC-V and builds on OpenBLAS kernels, extending it to additional architectures, combining layout propagation with low-level tuning, and integrating it more deeply with compiler-driven optimization frameworks are natural directions for future work. More broadly, LP-GEMM suggests a shift from optimizing isolated GEMM calls toward optimizing end-to-end workloads, with layout propagation beyond GEMM, such as for convolutions, representing a promising avenue for future research.

\bibliographystyle{IEEEtranDOI}
\bibliography{bibliografia}

\end{document}